# Diverse Group Formation Based on Multiple Demographic Features

Mohammed Alqahtani, Susan Gauch, Omar Salman, Mohammed Ibrahim, Reem Al-Saffar
*University of Arkansas*
*Department of Computer Science, Fayetteville, AR, US*
{ma063, sgauch, oasalman, msibrahi, rbalsaff}@uark.edu

Abstract: The goal of group formation is to build a team to accomplish a specific task. Algorithms are employed to improve the effectiveness of the team so formed and the efficiency of the group selection process. However, there is concern that team formation algorithms could be biased against minorities due to the algorithms themselves or the data on which they are trained. Hence, it is essential to build fair team formation systems that incorporate demographic information into the process of building the group. Although there has been extensive work on modeling individuals' expertise for expert recommendation and/or team formation, there has been relatively little prior work on modeling demographics and incorporating demographics into the group formation process.

We propose a novel method to represent experts' demographic profiles based on multidimensional demographic features. Moreover, we introduce two diversity ranking algorithms that form a group by considering demographic features along with the minimum required skills. Unlike many ranking algorithms that consider one Boolean demographic feature (e.g., gender or race), our diversity ranking algorithms consider multiple multivalued demographic attributes simultaneously. We evaluate our proposed algorithms using a real dataset based on members of a computer science program committee. The result shows that our algorithms form a program committee that is more diverse with an acceptable loss in utility.

## 1 INTRODUCTION

Different research areas have investigated the process of team formation with the goal of forming an innovative team. One key to a successful team is having qualified and collaborative team members who work as a team to achieve all tasks (Brocco, Hauptmann, & Andergassen-Soelva, 2011) (Lappas, Liu, & Terzi, 2009). Other systems build the team based on the social network in which they claimed that the quality of relationship could influence the success of the team (Lappas, Liu, & Terzi, 2009). Most of those systems focused on the expertise and the strength of the experts' relationship. In addition, there are several algorithms try to automate the process of recommending members to join the group, however, these methods can involve bias (Feldman, Friedler, Moeller, Scheidegger, & Venkatasubramanian, 2015) (Kamishima, Akaho, & Sakuma, 2011) (Zehlike, et al., 2017).

The issue of bias has been discovered in different areas in everyday life, industry, and academia. In academia, different studies investigated this issue and presented several features that considered as potential sources of bias. Those features are gender (Bornmann & Daniel, 2005) (Lerback & Hanson, 2017), ethnicity (Gabriel, 2017), geolocation (Murray, et al., 2019), career stage (Holman, Stuart-Fox, & Hauser, 2018) (Lerback & Hanson, 2017), and institution impact (Bornmann & Daniel, 2005). These biases can affect peer review, tenure and promotion, and career advancement. Some work has been done to address in the issue of bias by developing fair algorithms based on a single feature at a time, either gender or ethnicity (Zehlike, et al., 2017). However, in our study, we evaluate algorithms that incorporate multiple demographic features (gender, ethnicity, geolocation, career stage, and institution impact), that believed to be the main source of bias in academia.

In our research, we focus on the issue of bias in conference program committee (PC) formation. Because of its importance in career advancement, there have been several efforts to make the peer review process more transparent and less susceptible to various types of bias. One recommendation to reduce unfairness is to introduce diversity (Hunt, Layton, & Prince, 2015) amongst the reviewers that, in the case of a conference, begins with increasing the diversity of the PC. Increasing diversity can also enhance the work outcomes and produces a positive influence on scientific performance (AlShebli, Rahwan, & Woon, 2018).

It is clear that there is a lack of diversity within Computer Science as a whole. For example, fewer than 27% CS professionals are female and whites dominate more than 65% of CS professionals (Khan, Robbins, & Okrent, 2020). This lack of diversity is reflected in participation in conferences (Holman, Stuart-Fox, & Hauser, 2018) and in the members of the program committees that govern academic conferences in Computer Science (Lerback & Hanson, 2017). Addressing this, SIGCHI, one of the highest impact ACM conferences, announced an explicit goal to increase the diversity of their PC in 2020 (SIGCHI, 2019).

To this end, we study the problem of introducing diversity in the process of algorithmically forming a group. We introduce a demographic profile for the candidate experts based on the multiple features (gender, ethnicity, geolocation, career stage, and institution impact). Although our algorithms should be applicable to any team formation domain, we currently focus on conference program committee formation. We introduce and assess two approaches to selecting candidates to join a PC based on their demographic features, considering candidates whose paper has been accepted by the conference in previous years to have the minimum expertise necessary to join the PC. The main contributions of this paper are:

- Develop expert demographic profiles that consist of multiple features.
- Develop and assess algorithms to form a group based on diversity.

## 2 RELATED WORK

We begin by reviewing previous work in demographic user modeling and group formation approaches and then discuss several aspects related to the issue of bias.

### 2.1 Demographic Information

User profiles are an integral part of all work into personalization (Gauch, et al., 2007); one can't create frameworks that adjust to an individual without having an accurate model of the user's abilities and needs. Numerous investigations have demonstrated the importance of incorporating demographic features when developing automated frameworks to select choices for individuals (Khalid, Salim, Loke, & Khalid, 2011). However, online profiles ordinarily choose to not collect this data since clients are frequently worried about how such data might be utilized. Thus, in order to utilize demographic information to ensure fairness and anti-discrimination in their algorithms, organizations often, infer features such as gender, nationality, and ethnicity based on the user's name (Chandrasekaran, Gauch, Lakkaraju, & Luong, 2008).

The determination of which demographic features to include varies from one environment to another. In academia, for instance, demographic profiling typically considers features such as ethnicity, age, gender, race, and socioeconomic background (Cochran-Smith & Zeichner, 2009). There have been several approaches to extract demographic information, specifically gender and ethnicity (Dias & Borges, 2017) (Michael, 2007). However, in our research, we use an NamSor API used by Jain et al (Jain & Minni, 2017) to extract gender based on the user's names. This tool covers more than 142 languages and the overall gender precision and recall of this tool are respectively 98.41% and 99.28% (blog, NamSor, 2018).

### 2.2 Group Formation

One key to a successful organization is having a good leader and collaborative group who work as a team to achieve all tasks (Wi, Oh, Mun, & Jung, 2009). Based on this insight, several automatic team formation methods have been proposed to form groups based on a social network (Lappas, Liu, & Terzi, 2009) (Owens, Mannix, & Neale, 1998). (Juang, Huang, & Huang, 2013) (Wi, Oh, Mun, & Jung, 2009) (Brocco, Hauptmann, & Andergassen-Soelva, 2011). However, considering social network relationships may result in bias. To address this, Chen et al. (Chen, Fan, Ma, & Zeng, 2011) proposed a genetic grouping algorithm to automatically construct a group of reviewers that balances inclusion based on age, region, or professional title. We take a similar approach in our study; however, we form a group of candidates with respect to five demographic features simultaneously.

Several projects on group formation have focused specifically on academia. For instance, Wang et al. (Wang, Lin, & Sun, 2007) introduced DIANA algorithms that consider several parameters to build a group of students. Tabo et al. (Tobar & de Freitas, 2007) proposed a method to create a team for academic duties within a class. Although these methods automate the procedure of recommending a candidate to be a member of the team, because they do not specifically incorporate demographic modeling, those approaches may lead to bias.

## 2.2 Fairness

Fairness necessitates that underrepresented groups should have the same access to opportunities as the population as a whole (Zehlike, et al., 2017). Hence, (Feldman, Friedler, Moeller, Scheidegger, & Venkatasubramanian, 2015) investigated the problem of unintentional bias and how it impacts various populations that should be treated similarly.

*Fairness in Machine Learning.* With the increased use of machine learning in many aspects of everyday life, there is increasing concern that these systems make decisions in an unbiased way (Asudeh, Jagadish, Stoyanovich, & Das, 2019) (Feldman, Friedler, Moeller, Scheidegger, & Venkatasubramanian, 2015). Several investigations have demonstrated that, although classifiers themselves are generally not biased, the outcome of those classifiers may be affected by the bias in the training data (Feldman, Friedler, Moeller, Scheidegger, & Venkatasubramanian, 2015) (Kamishima, Akaho, & Sakuma, 2011) (Asudeh, Jagadish, Stoyanovich, & Das, 2019). In response, (Zemel, Wu, Swersky, Pitassi, & Dwork, 2013) derived a learning algorithm for fair classification by providing suitable data representation and at the same time obfuscating any data about membership in a protected group.

*Bias in Academia.* The issue of bias in academia has been well studied. Gabriel (Gabriel, 2017) presents a study that demonstrates that ethnicity discrimination still exists in British academia. As an example, black professors represent only 0.1% of all professors in the UK although they constitute up to 1.45% of the UK population. Bornmann et al (Bornmann & Daniel, 2005) investigated the impact of bias on the process of selecting doctoral and post-doctoral members. They found evidence of bias based on gender, area of research, and affiliation, but not nationality.

*Bias in Peer Review.* The peer review process is one of the most-studied areas of research into bias in academia. Lee et al. (Lee, Sugimoto, Zhang, & Cronin, 2013) studied different kinds of bias in peer review and how it impacts the review process of accepting or rejecting submitted articles. (Holman, Stuart-Fox, & Hauser, 2018) and (Lerback & Hanson, 2017) provided evidence that females are persistently underrepresented in publications from computer science, math, physics, and surgery. This was further confirmed by (Murray, et al., 2019) They found evidence that a reviewer is more likely to accept publications by authors of the same gender and from the same country as themselves. Hence, many publications and conferences have adopted a double-blind review to avoid this type of bias. However, several studies show that 25–40% of the time, reviewers can recognize authors (Baggs, Broome, Dougherty, Freda, & Kearney, 2008) (Justice, Cho, Winker, & Berlin, 1998), which can lead to bias. Lane (Lane D. , 2008) suggest that within specific fields, these numbers could be higher.

In order to address the issue of bias in academia, Yin et al. (Yin, Cui, & Huang, 2011) studied the relationships between bias and three features: the reviewer's reputation, the co-authorship connection, and the coverage. They suggested that to avoid biased results, one should ensure diversity in the peer review committee itself. Other studies by (Wang, et al., 2016) (Lane T. , 2018) (Chen, Fan, Ma, & Zeng, 2011) suggested that increasing the diversity in a peer review committee will enhance the review process and lead to better outcomes.

To summarize, bias exists within academia even though the research community has taken strides to avoid it. Several studies indicate that increasing diversity when forming a group can enhance its quality of work and produce fairer results. Most current approaches concentrate on a single protected feature at a time, e.g., gender or race. However, in our research, we contribute to this research area by developing algorithms that consider multiple features simultaneously.

## 3 DEMOGRAPHIC PROFILE MODELING

In this section, we present how we collect our demographic data collection process (3.1). Then, we describe how we determine the protected groups and the procedure of mapping our demographic data to Boolean weighted features (3.2).

## 3.1 Data Collection

Our demographic profile consists of five features that have been identified as potential sources of bias in academia, to whit Gender, Ethnicity, Geolocation,

University Rank, and Career Stage. For each researcher in our pool of PC candidates we use publicly accessible information to collect their demographic data. We collected this information using web-scraping scripts that we developed to automate the process. The following explains our method of collecting each of the demographic feature values: **Gender:** We determine the gender of each scholar using NamSor (NamSor, 2020), a tool that predicts gender based on an individual's full name. It also returns the degree of confidence in the prediction in a range between 0 and 1 NamSor gender API has an overall 98.41% precision and 99.28% recall (blog, NamSor, 2018). Additionally, the NamSor inventor used official directory of the European union (Union, European, 2020) to assess the NamSor API gender for European names. The gender error rate was only less than 1%. **Ethnicity:** We determine the ethnicity of each scholar using NamSor (NamSor, 2020), a tool used to extract the ethnicity of an individual based on that individual's full name. For each name, they return the most likely ethnicity from a set of 5 possibilities, e.g., W_NL (for White), or B_NL (for Black). NamSor is widely used to predict ethnicity and gender in other studies **Geolocation:** The location is obtained using a scholar profile in Google Scholar (Google Scholar, 2020). We extract the university name at which each scholar works. Then, we use that information to locate the university's home page and, from that, determine the country in which they work. Additionally, for those in the United States, we also extract the state in which they work. **University Rank:** Using the university name extracted above, we use the Times Higher Education (Education, Times Higher, 2020) to determine the university's rank. This site produces ranks for each institution between 1 to 1400, so we use the value 1401 for unranked universities. **Career Stage:** We extract the academic position of each scholar using their profile in Google Scholar (Google Scholar, 2020). **H-index:** We also collect the h-index from the scholar's Google Scholar profile (Google Scholar, 2020) and use this feature to measure the utility of the various PCs. **Note:** Candidate PC members without a Google Scholar profile, and those without academic positions, are omitted from our dataset.

## 3.2 Mapping Boolean Weights

Factors such as culture and environment may affect the definition of protected groups, (Feldman, Friedler, Moeller, Scheidegger, & Venkatasubramanian, 2015). However, our definition of protected groups is based on which group is underrepresented in the population being studied, i.e., researchers in Computer Science. Each feature in a scholar's profile is represented using a Boolean weight, typically 1 if the scholar is a member of the protected (underrepresented) group and 0 otherwise. The following illustrates how we determine the values of each feature:

*Gender.* Females make up 27% of professionals in Computer Science in 2017 (Khan, Robbins, & Okrent, 2020), so they are the protected group.

*Ethnicity:* In computer science and engineering, whites make up the majority of professionals at 65% (Khan, Robbins, & Okrent, 2020), so non-white ethnicities other ethnicities are considered the protected group.

*Geolocation*: In this feature, we utilize the GDP (Gross Domestic Production) retrieved from World Development Indicators database (Worldbank, 2018) to divide the countries into *developing* or *developed*. We compute the average world GDP and then employ this to partition the values of this feature into a developing country for those who below the average and developed country otherwise. The developing country is our protected group. For those who live in the United States, we use the Established Program to Stimulate Competitive Research (EPSCoR) designation (Foundation, National Science, 2019) developed by the NSF (National Science Foundation). EPSCoR states, those with less federal grant funding, are the protected group.

*University Rank*: Based on the rankings provided by (Education, Times Higher, 2020), we use the mean to partition university ranks into low-ranked and high-ranked groups and use low-ranked universities (higher values) as the protected group.

*Career Stage:* We consider tenured faculty, those who are associated professor or higher as senior; otherwise they are considered junior and consider., junior researchers as our protected group.

---

**Algorithm 1** Univariate Greedy

1. *priority_queue ← Initialize an empty queue*
2. *For each profile:*
3.     *Diversity score ← compute profile score*
4.     *Add profile to priority_queue using diversity score as priority order*
5. *candidates ← Select N profiles from top of priority_queue*

---

In summary, each demographic profile consists of five features (gender, ethnicity, geolocation, university rank, and career stage) associated with a Boolean weight that represents whether or not the candidate is a member of the protected group for that feature. We also collect each researcher's h-index from their Google Scholar profile that we use to evaluate the utility of each PC in our evaluation.

## 4 METHODOLOGY

In this section, we begin to introduce our fair group formation algorithms.

### 4.1 Univariate Greedy Algorithm

In section (3), we described our demographic profiles and the process of mapping all values to Boolean weights. Based on that, we can compute each candidate's diversity score ($Score_D$) by summing the weights for each demographic feature $d_i$ as shown in equation (1).

$$Score_D = \sum_{i=1}^{n} d_i \qquad (1)$$

Once we obtain the diversity score for each candidate, we apply our Univariate Greedy Algorithm (UGA) that selects candidates to join the group. To accomplish this process, we place the candidates into a priority queue based on their diversity score. Then, we iteratively remove the top candidate from the priority queue until the targeted group size achieved. For instance, when forming a program committee for a conference, the desired PC size is set to the size of the current, actual PC for that conference. If two or more candidates have the same diversity score, we select one of those candidates randomly.

### 4.2 Multivariate Greedy Algorithm

The previous method maximizes the diversity score of the resulting group, but it does not guarantee multidimensional diversity among the resulting group members. It could result in a high diversity score by selecting an entirely female group, for example, while accidentally excluding any members from ethnic minority groups. Thus, we developed a Multivariate Greedy Algorithm (MGA) to address this issue by creating one priority queue per demographic feature and using a round robin algorithm to select a member from each queue until the group size is achieved. In particular, we build five priority queues, one per feature in our current demographic profile, each of which contains a list of all candidates sorted based on that feature. Round robin selection is used to select the highest-ranked unselected candidate from each queue in turn. Once a candidate is selected, it is removed from all queues to avoid choosing the same researcher repeatedly. This process continues iteratively until the group is formed. Note: currently the weights for the features are just 1 or 0; in the case of a tie, one candidate is selected randomly. In future, we will implement and evaluate non-Boolean feature weights.

Table 1: Composition of our datasets.

| Dataset | PC Members | Authors | Total |
|---------|------------|---------|-------|
| SIGCHI17 | 213 | 436 | 649 |
| SIGMOD17 | 130 | 290 | 420 |
| SIGCOMM17 | 23 | 125 | 148 |

## 5 EXPERIMENT AND RESULT

We now introduce our dataset and describe the process of evaluating our algorithms.

### 5.1 Datasets

For our driving problem, we want to focus on the PC members for high impact computer science conference. Thus, we are building a dataset based on ACM conferences and we select a conference based on several criteria: 1) the conferences should have high impact; 2) the conferences should have little or no overlap in topics; 3) the conferences should have a reasonably large number of PC members and accepted papers. Based on these criteria, we selected SIGCHI (The ACM Conference on Human Factors in Computing Systems), SIGMOD (Symposium on Principles of Database Systems), and SIGCOMM (The ACM Conference on Data Communication. We evaluate our diverse group formation algorithms using subsets of datasets that consists of the PC and authors of all accepted papers of the three selected conferences. We exclude candidates who: 1) do not have a Google Scholar profile; 2) are missing at least one feature's value; 3) primarily worked in the industry. Based on these criteria, we create a pool for each conference that contains both PC members and authors of accepted papers (see Table 1). The demographic distribution of those PC's is summarized in Figures 1 and 2. These clearly illustrate that all of the three PC's had a low participation rate from all protected groups. As an example, SIGCOMM 2017 had only 8.7% female PC members and, SIGMOD 2017's PC was only 17.7% female. Similarly, whites dominate with 78.40% of SIGCHI 2017 PC, 55.38% of SIGMOD 2017 PC, and 69.56% of SIGCOMM 2017 PC.

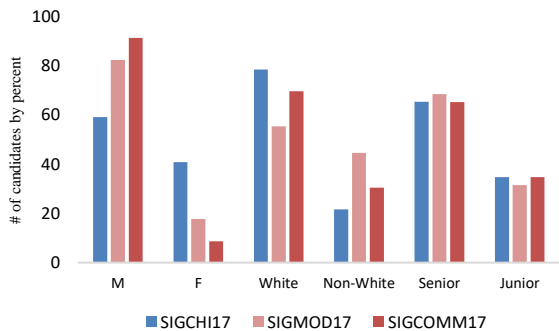

Figure 1: Data Distribution of the three current PC's for Gender, Race, and Career Stage Features.

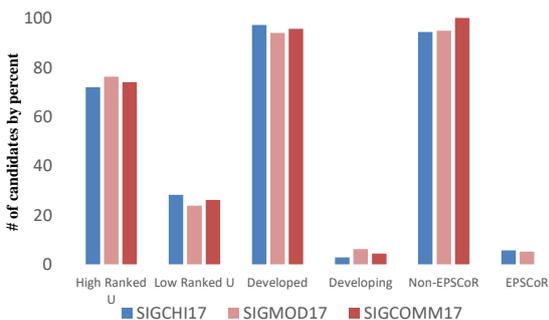

Figure 2: Data Distribution of the three current PC's for Affiliation Impact, and Geolocation Features.

## 5.2 Baseline and Metrics

*Baseline.* Our baseline is a Random Selection Algorithm (RSA) that randomly selects candidates to form the group without considering diversity.
*Metrics.* Our algorithms attempt to generate a more diverse PC. We evaluate their effectiveness using *Diversity Gain* ($D_G$) of our proposed PCs versus the baseline:

$$D_G = \text{MIN}\left(100, \frac{\sum_{i=1}^{n} \rho_{G_i}}{n}\right) \quad (2)$$

where $\rho_{G_i}$ is the relative percentage gain for each feature versus the baseline, divided by the total number of features $n$. Each feature's diversity gain is capped at a maximum value of 100 to prevent a large gain in a single feature dominating the value.

---
**Algorithm 2** Multivariate Greedy
---
1. *feature_names* ← List of all queue names, one per features
2. For each *feature* in *feature_names*:
3.     *priority_queue*[*feature*] ← Initialize an empty queue
4. For each *profile*:
5.     For each *feature* in *feature_names*:
6.         *score*[*feature*] ← compute *profile* score for each *feature*
7.         Add *profile* to [*feature*] using *score*[*feature*] as priority order
8. *candidates* ← empty list
9. While number of *candidates* < *N*:
10.     *feature* ← *feature_names*[0]
11.     Repeat:
12.         *candidate* ← Get and remove profile from *priority_queue*[*feature*]
13.         Until *candidate* is not in *candidates*
    Add *candidate* to *candidates*
14.     Rotate *feature* to end of *feature_names*.
15. Now we have *N* candidates selected.
---

By choosing to maximize diversity, it is likely that the expertise of the resulting PC will have slightly lower expertise. To measure this drop in utility, we use the average h-index of the PC members and compute the utility loss ($UL_i$) for each proposed PC using the following formula:

$$UL_i = U_b - U_{P_j} \quad (3)$$

where $U_{P_j}$ is the utility of PC$i$ and $U_b$ is the utility of the baseline. We then compute the utility savings ($\Upsilon_i$) of PC$i$ relative to the baseline as follows:

$$\Upsilon_i = \frac{UL_i}{U_b} \quad (4)$$

Finally, we compute the F measure (Jardine, 1971) to examine the ability of our algorithms to balance diversity gain and utility savings:

$$F = 2 * \frac{D_G * \Upsilon_i}{D_G + \Upsilon_i} \quad (5)$$

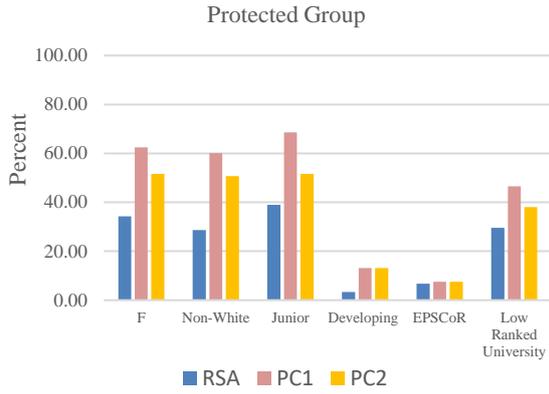

Figure 6: Comparison of the protected groups improvement between the baseline PC produced by the baseline and proposed PCs of SIGCHI 2017 produced by our UGA and MGA.

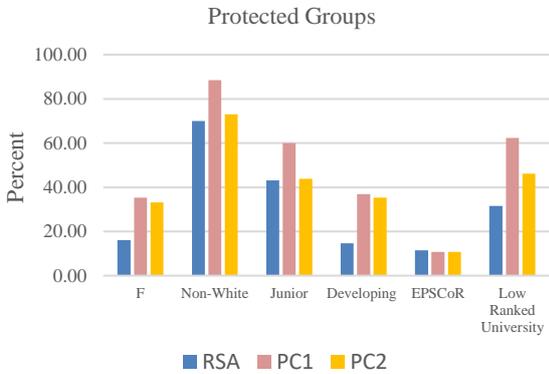

Figure 7: Comparison of the protected groups improvement between the PC produced by the baseline and proposed PCs of SIGMOD 2017 produced by our UGA and MGA.

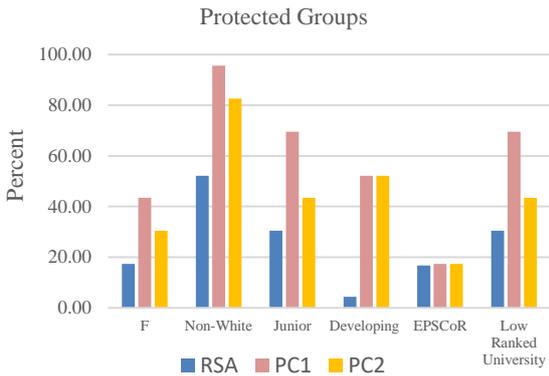

Figure 8: Comparison of the protected groups improvement between the PC produced by the baseline and proposed PCs of SIGCOMM 2017 produced by our UGA and MGA.

## 5.3 Results

*Comparison with the baseline*. Our algorithms produce ranked list(s) from which we select to form the PCs with the overarching goal of increasing the diversity in the program committee. Hence, we report the differences between the PC produced by the baseline, random selection (RSA), and the PCs proposed by the algorithms described in Section 4. Looking at Figures 6, 7, and 8, we can see that both algorithms succeeded in increasing the diversity in the recommended PCs in all demographic groups except EPSCoR. In some cases, the UGA overcorrects and, in its efforts to select diverse members, we end up with a demographically biased PC in favor of some protected groups, e.g., female. We must also compare the effect of the algorithms with respect to the expertise of the resulting PC. Table 2 summarizes the diversity gain (DG), utility loss (UL), utility Savings ($\Upsilon_i$), and F scores for the PCs proposed by each algorithm. The MGA and UGA obtained diversity gains of over 46 for all of the three proposed PC's, with the biggest gain occurring for SIGCHI2017 using the UGA. The gains in diversity occur with an average utility loss of 35.24% for the UGA but only 15.41% for the MGA. Since the MGA resulted in greater utility savings, its average F score is higher, and we conclude that the Multivariate Greedy Algorithm outperforms the Univariate Greedy Algorithm.

Table 2: Experimental results for the UGA and MGA algorithms versus the RSA (baseline). All values presented as percentages.

| Table | $D_G$ | $UL_i$ | $\Upsilon_i$ | F |
|---|---|---|---|---|
| SIGCHI | | | | |
| UGA | **67.18** | 32.88 | 67.12 | 67.15 |
| MGA | 55.5 | 20.67 | 79.33 | 65.31 |
| SIGMOD | | | | |
| UGA | 50.51 | 17.47 | 82.53 | 62.67 |
| MGA | 53.00 | **-2.42** | 102.42 | 69.85 |
| SIGCOMM | | | | |
| UGA | 46.80 | 55.37 | 44.63 | 45.69 |
| MGA | 50.56 | 27.99 | 72.01 | 59.41 |
| Average | | | | |
| UGA | **54.83** | 35.24 | 64.76 | 58.50 |
| MGA | 53.02 | **15.41** | 84.59 | 64.86 |

Table 3: Comparison of all the three proposed PC's produced by our MGA and the current PC's. All values presented as percentages.

| Feature | Method | SIGCHI 2017 | SIGMOD 2017 | SIGCOMM 2017 | Average |
|---|---|---|---|---|---|
| Female | Current | 40.85 | 17.69 | 8.7 | 22.41 |
| | MGA | 48.83 | 29.23 | 30.43 | 36.16 |
| Non-White | Current | 21.6 | 44.62 | 30.43 | 32.22 |
| | MGA | 52.11 | 78.46 | 69.57 | 66.71 |
| Junior | Current | 34.74 | 31.54 | 34.78 | 33.69 |
| | MGA | 48.83 | 49.23 | 52.17 | 50.08 |
| Developing | Current | 2.81 | 6.15 | 4.35 | 4.44 |
| | MGA | 14.08 | 34.62 | 52.17 | 33.62 |
| EPSCoR | Current | 5.71 | 2.31 | 0.00 | 2.67 |
| | MGA | 7.51 | 10.77 | 17.39 | 11.89 |
| Low Rank University | Current | 28.17 | 23.85 | 26.09 | 26.04 |
| | MGA | 49.30 | 49.23 | 47.83 | 48.79 |

Table 4: Comparison of the average h-index of each proposed PC produced by our MGA versus the current PCs.

| Table | Current | MGA |
|---|---|---|
| SIGCHI2017 | 24.55 | 19.04 |
| SIGMOD2017 | 32.86 | 23.27 |
| SIGCOMM2017 | 29.43 | 19.91 |
| Average | 28.95 | 20.74 |

### 5.4 Validation

Finally, we provide a comparison between the actual PC's for the three conferences and the PCs proposed by our best algorithm, the MGA (see Table 3).

The number of PC members from the protected groups were increased across all demographic features for all conferences. In most cases the algorithm did not over-correct by including more than 50% of any protected demographic group, with the exception of the participation of non-white that was increased to over 66.7%. The participation of females and junior researchers all increased about 50% and non-whites and researchers from lower-ranked universities doubled. Researchers from the developing world and EPSCoR states increased many-fold, although this was achieved by selecting all candidates from EPSCoR states and most candidates from developing countries. The h-index for the proposed PC dropped 28.35% (see Table 4). The overall diversity gain for the proposed PC is 53.02%, the utility savings 84.59% and the F-measure 64.86%.

## 6 CONCLUSION AND FUTURE WORK

Groups of experts are formed in many situations within industry and academia. However, there may be bias in the traditional group formation process leading to inferior results and blocking members of underrepresented populations from access to valuable opportunities. We investigate the issue of bias in academia, particularly the formation of conference program committees, and develop algorithms to form a diverse group of experts. Our approach is based on representing candidate experts with a profile that models their demographic information consisting of five features that might be sources of bias, i.e., gender, ethnicity, career stage, geolocation, and affiliation impact. Most previous work focuses on algorithms that guarantee fairness based on a single, Boolean feature, e.g., race, gender, or disability. We consider five Boolean features simultaneously and evaluated two group formation algorithms. The Univariate Group Algorithm (UGA) selects members based on a composite diversity score and the Multivariate Group Algorithm (MGA) selects members based on a round robin of priority queues for each diversity feature. The resulting proposed PCs were compared in terms of diversity gain and utility savings, as measured by a decrease in the average h-index of the PC members. The MGA produced the best results with an average increase of 48.42% per protected group with utility loss of only 10.21% relative to a random selection algorithm.

In some cases, our algorithms overcorrected, producing a PC that had overrepresentation from protected groups. In future, we will develop new algorithms that have demographic parity as a goal so that the PC composition matches the demographic distributions in the pool of candidates. These will require modifications to our MGA so that the feature queues are visited proportionally to the protected group participation in the pool. We will also explore the use of non-Boolean feature weights and dynamic algorithms that adjust as members are added to the PC.

In conclusion, our proposed work provides new ways to create inclusive, diverse groups to provide better opportunities, and better outcomes, for all.

# REFERENCES

AlShebli, B. K., Rahwan, T., & Woon, W. L. (2018). Ethnic diversity increases scientific impact. *arXiv preprint arXiv:1803.02282*.

Asudeh, A., Jagadish, H. V., Stoyanovich, J., & Das, G. (2019). Designing fair ranking schemes. *In Proceedings of the 2019 International Conference on Management of Data (pp. 1259-1276). ACM.*

Baggs, H., Broome, M., Dougherty, M., Freda, M., & Kearney, M. (2008). Blinding in peer review: the preferences of reviewers for nursing journals. *Journal of Advanced Nursing, 64(2), 131–138.*

blog, NamSor. (2018, Jan 31). *Inferring The World's Gender and Ethnic Diversity using Personal Names.* Retrieved 2020., from https://namesorts.com/2018/01/31/understanding-namsor-api-precision-for-gender-inference/

Bornmann, L., & Daniel, H. D. (2005). Selection of research fellowship recipients by committee peer review. Reliability, fairness and predictive validity of Board of Trustees' decisions. *Scientometrics, 63(2), 297-320.*

Brocco, M., Hauptmann, C., & Andergassen-Soelva, E. (2011). Recommender system augmentation of HR databases for team recommendation. *Paper presented at the Database and Expert Systems Applications (DEXA), 22nd International Workshop On, 554-558.*

Chandrasekaran, K., Gauch, S., Lakkaraju, P., & Luong, H. P. (2008). Concept-based document recommendations for citeseer authors. *International Conference on Adaptive Hypermedia and Adaptive Web-Based Systems.* Springer, Berlin.

Chen, Y., Fan, Z. P., Ma, J., & Zeng, S. (2011). A hybrid grouping genetic algorithm for reviewer group construction problem. *Expert Systems with Applications, 38(3), 2401-2411.*

Cochran-Smith, M., & Zeichner, K. M. (2009). Studying teacher education: The report of the AERA panel on research and teacher education. *Routledge.*

Dias, T. G., & Borges, J. (2017). A new algorithm to create balanced teams promoting more Diversity. *European Journal of Engineering Education, 42(6), 1365-1377. doi:10.1080/03043797.2017.1296411.*

Education, Times Higher. (2020). *Times.* Retrieved from https://www.timeshighereducation.com/

Feldman, M., Friedler, A., Moeller, J., Scheidegger, C., & Venkatasubramanian, S. (2015). Certifying and removing disparate impact. *In Proceedings of the 21th ACM SIGKDD International Conference on Knowledge Discovery and Data Mining (pp. 259-2).*

Foundation, National Science. (2019). *EPSCoR states.* Retrieved from https://www.nsf.gov/od/oia/programs/epscor/nsf_oiia_epscor_EPSCoRstatewebsites.jsp

Gabriel, D. (2017). Race, racism and resistance in British academia. *In Rassismuskritik und Widerstandsformen (pp. 493-505). Springer VS, Wiesbaden.*

Gauch, S., Speretta, M., Chandramouli, A., Micarelli, A., Brusilovsky, P., Kobsa, A., & Nejdl, W. (2007). The adaptive Web: methods and strategies of Web personalization.

Google Scholar. (2020). *Google.* Retrieved 2020, from https://scholar.google.com/

Holman, L., Stuart-Fox, D., & Hauser, C. E. (2018). The gender gap in science: How long until women are equally represented?. *PLoS biology, 16(4), e2004956.*

Hunt, V., Layton, D., & Prince, S. (2015). Diversity matters. *McKinsey & Company, 1, 15-29.*

Jain, A., & Minni, J. (2017). Location based Twitter Opinion Mining using Common-Sense

Information. *Global Journal of Enterprise Information System, 9(2)*.

Jardine, N. &. (1971). The use of hierarchic clustering in information retrieval. . *Information storage and retrieval*, (pp. 7(5), 217-240.).

Juang, M. C., Huang, C. C., & Huang, J. L. (2013). Efficient algorithms for team formation with a leader in social networks. *The Journal of Supercomputing, 66(2), 721-737*.

Justice, A., Cho, M., Winker, M., & Berlin, J. (1998). Does masking author identity improve peer review quality? A randomized controlled trial. *Journal of the American Medical Association, 280(3), 240–242*.

Kamishima, T., Akaho, S., & Sakuma, J. (2011). Fairness-aware learning through regularization approach. *In 2011 IEEE 11th International Conference on Data Mining Workshops (pp. 643-650). IEEE*.

Khalid, K., Salim, H. M., Loke, S. P., & Khalid, K. (2011). Demographic profiling on job satisfaction in Malaysian utility sector. *International Journal of Academic Research, 3(4), 192-198*.

Khan, B., Robbins, C., & Okrent, A. (2020, Jan 15). *Science and Engineering Indicator*. Retrieved from https://ncses.nsf.gov/pubs/nsb20198/demographic-trends-of-the-s-e-workforce

Lane, D. (2008). Double-blind review: Easy to guess in specialist fields. *Nature, 452, 28*.

Lane, T. (2018, 10 12). *Diversity in Peer Review: Survey Results*. (COPE) Retrieved from https://publicationethics.org/news/diversity-peer-review-survey-results

Lappas, T., Liu, K., & Terzi, E. (2009). Finding a team of experts in social networks. *In the Proceedings of the 15th ACM SIGKDD International Conference on Knowledge Discovery and Data Mining, 467-476.*, (pp. Lappas, T.; Liu, K.; Terzi, E.).

Lee, C. J., Sugimoto, C. R., Zhang, G., & Cronin, B. (2013). Bias in peer review. *Journal of the American Society for Information Science and Technology, 64(1), 2-17*.

Lerback, J., & Hanson, B. (2017). Journals invite too few women to referee. *Nature News, 541(7638), 455*.

Michael, J. (2007). 40000 namen, anredebestimmung anhand des vornamens. *C'T, 182-183*.

Murray, D., Siler, K., Lariviére, V., Chan, W. M., Collings, A. M., Raymond, J., & Sugimoto, C. R. (2019). Gender and international diversity improves equity in peer review. *BioRxiv, 400515*.

NamSor. (2020). *NamSor*. Retrieved 2020, from https://www.namsor.com/

Owens, D. A., Mannix, E. A., & Neale, M. A. (1998). Strategic formation of groups: Issues in task performance and team member selection. *Research on managing groups and teams, 1(1998), 149-165*.

Rodriguez, M. A., & Bollen, J. (2008). An algorithm to determine peer-reviewers. *In Proceedings of the 17th ACM conference on Information and knowledge management (pp. 319-328). ACM*.

Salman, O., Gauch, S., Alqahtani, M., & Ibrahim, M. (2020). The Demographic Gap in Conference Committees.

SIGCHI. (2019). *Diversity of the Program Committee for CHI 2020*. Retrieved from https://chi2020.acm.org/blog/diversity-of-the-program-committee-for-chi-2020/

Singh, A., & Joachims, T. (2018). Fairness of exposure in rankings. *In Proceedings of the 24th ACM SIGKDD International Conference on Knowledge Discovery & Data Mining (pp. 2219-2228). ACM*.

Tobar, C. M., & de Freitas, R. L. (2007). A support tool for student group definition. *In 2007 37th Annual Frontiers In Education Conference-Global Engineering: Knowledge Without Borders, Opportunities Without Passports (pp. T3J-7). IEEE*.

Union, European. (2020). *The official directory of the European Union*. Retrieved 2020, from https://op.europa.eu/en/web/who-is-who

Wang, D. Y., Lin, S. S., & Sun, C. T. (2007). DIANA: A computer-supported heterogeneous grouping system for teachers to conduct successful small learning groups. *Computers in Human Behavior, 23(4), 1997-2010*.

Wang, W., Kong, X., Zhang, J., Chen, Z., Xia, F., & Wang, X. (2016). Editorial behaviors in peer review. *SpringerPlus, 5(1), 903*.

Wi, H., Oh, S., Mun, J., & Jung, M. (2009). A team formation model based on knowledge and collaboration. *Expert Systems with Applications, 36(5), 9121-9134*.

Worldbank. (2018). *gdp-ranking*. Retrieved 2020, from https://datacatalog.worldbank.org/dataset/gdp-ranking

Yin, H., Cui, B., & Huang, Y. (2011). Finding a wise group of experts in social networks. *In International Conference on Advanced Data Mining and Applications (pp. 381-394)*. Springer, Berlin, Heidelberg.

Zehlike, M., Bonchi, F., Castillo, C., Hajian, S., Megahed, M., & Baeza-Yates, R. (2017). Fa* ir: A fair top-k ranking algorithm. *In Proceedings of the 2017 ACM on Conference on Information and Knowledge Management (pp. 1569-1578). ACM*.

Zemel, R., Wu, Y., Swersky, K., Pitassi, T., & Dwork, C. (2013). Learning fair representations. *In International Conference on Machine Learning*, (pp. (pp. 325-333).).